\documentclass[12pt]{article}
\usepackage{epsfig,amssymb,amsmath,psfrag,multirow,color,cite}
\numberwithin{equation}{section}

\newcommand{\be}{\begin{equation}}
\newcommand{\ee}{\end{equation}}
\newcommand{\bea}{\begin{eqnarray}}
\newcommand{\eea}{\end{eqnarray}}

\def\cN{{\mathcal N}}
\def\cO{{\mathcal O}}
\def\cL{{\mathcal L}}
\def\ws{{w^\ast}}

\catcode`\@=11
\font\manfnt=manfnt
\def\Watchout{\@ifnextchar [{\W@tchout}{\W@tchout[1]}}
\def\W@tchout[#1]{{\manfnt\@tempcnta#1\relax%
  \@whilenum\@tempcnta>\z@\do{%
    \char"7F\hskip 0.3em\advance\@tempcnta\m@ne}}}
\let\foo\W@tchout
\def\dubious{\@ifnextchar[{\@dubious}{\@dubious[1]}}

\def\@dubious[#1]{%
  \setbox\@tempboxa\hbox{\@W@tchout#1}
  \@tempdima\wd\@tempboxa
  \list{}{\leftmargin\@tempdima}\item[\hbox to 0pt{\hss\@W@tchout#1}]}
\def\@W@tchout#1{\W@tchout[#1]}
\catcode`\@=12
\font\cyr=wncyr8
\newcommand{\sha}{{\mbox{\cyr X}}}
\newfont{\scyr}{wncyr10 scaled 550}
\newcommand{\ssha}{\mbox{\bf \scyr X}}
\newcommand{\cZ}{{\cal Z}}

\def\beq{\begin{equation}}
\def\eeq{\end{equation}}

\def\bsp#1\esp{\begin{split}#1\end{split}}

\newcommand{\Hb}[0]{{\overline{H}}}

\newenvironment{sloppyequation}[0]{\normalsize\sloppy\begin{flushleft}\hspace*{0.75cm}\(\displaystyle}{\)\end{flushleft}\fussy\normalsize}

\newcommand{\beqsloppy}{\begin{sloppyequation}}
\newcommand{\eeqsloppy}{\end{sloppyequation}}

\begin{document}

\thispagestyle{empty}

\begin{flushright}
SLAC--PUB--15251
\end{flushright}

\begingroup\centering
{\Large\bfseries\mathversion{bold}
The six-point remainder function\\to all loop orders in the multi-Regge limit\par}%
\vspace{8mm}

\begingroup\scshape\large
Jeffrey Pennington\\
\endgroup
\vspace{6mm}
\begingroup\small
\emph{SLAC National Accelerator Laboratory, 
Stanford University, \\
Stanford, CA 94309, USA\\
Email:} {\tt jpennin@stanford.edu}\endgroup
\vspace{1.2cm}

\textbf{Abstract}\vspace{5mm}\par
\begin{minipage}{14.7cm}
We present an all-orders formula for the six-point amplitude of planar maximally supersymmetric $\cN=4$ Yang-Mills theory in the leading-logarithmic approximation of multi-Regge kinematics. In the MHV helicity configuration, our results agree with an integral formula of Lipatov and Prygarin through at least 14 loops. A differential equation linking the MHV and NMHV helicity configurations has a natural action in the space of functions relevant to this problem---the single-valued harmonic polylogarithms introduced by Brown. These functions depend on a single complex variable and its conjugate, $w$ and $\ws$, which are quadratically related to the original kinematic variables. We investigate the all-orders formula in the near-collinear limit, which is approached as $|w|\to 0$. Up to power-suppressed terms, the resulting expansion may be organized by powers of $\log|w|$. The leading term of this expansion agrees with the all-orders double-leading-logarithmic approximation of Bartels, Lipatov, and Prygarin. The explicit form for the sub-leading powers of $\log|w|$ is given in terms of modified Bessel functions.
\end{minipage}\par
\endgroup

\newpage
\section{Introduction}

In recent years, considerable progress has been made in the study of relativistic scattering amplitudes
in gauge theory and gravity. A growing set of computational tools, including unitarity \cite{UnitarityMethod}, BCFW recursion \cite{Britto2004ap,Britto2005fq,ArkaniHamed2010kv,ArkaniHamed2010gh}, BCJ duality \cite{Bern2008qj,Bern2010ue}, and symbology \cite{symbolsC,symbolsB,symbols,Goncharov2010jf,Duhr2011zq}, has facilitated many impressive perturbative calculations at weak coupling. The AdS/CFT correspondence has provided access to the new, previously inaccessible frontier of strong coupling~\cite{Alday2007hr}. The theory that has reaped the most benefit from these advances is, arguably, maximally supersymmetric $\cN=4$ Yang-Mills theory, specifically in the planar limit of a large number of colors. Indeed, $\cN=4$ super-Yang-Mills theory provides an excellent laboratory for the AdS/CFT correspondence, as well as for the structure of gauge theory amplitudes in general.

One of the reasons for the relative simplicity of $\cN=4$ super-Yang-Mills theory is its high degree of symmetry. The extended supersymmetry puts strong constraints on the form of scattering amplitudes, and it guarantees a conformal symmetry in position space. Recently, an additional conformal symmetry was found in the planar theory~\cite{Alday2007hr,Drummond2006rz,Bern2006ew,Drummond2007aua,Brandhuber2007yx,Alday2007he,Drummond2007au}. It acts on a set of dual variables, $x_i$, which are related to the external momenta $k_i^\mu$ by $k_i=x_i-x_{i+1}$. At tree level, this dual conformal symmetry can be extended to a dual super-conformal symmetry~\cite{Drummond2008vq} and even combined with the original conformal symmetry into an infinite-dimensional Yangian symmetry~\cite{Drummond2009fd}. At loop level, the dual conformal symmetry is broken by infrared divergences. According to the Wilson-loop/amplitude duality~\cite{Alday2007hr,Drummond2007aua,Brandhuber2007yx}, these infrared divergences can be understood as ultraviolet divergences of particular polygonal Wilson loops. In this context, the breaking of dual conformal symmetry is governed by an anomalous Ward identity~\cite{Drummond2007au,Drummond2007cf}. For maximally-helicity violating (MHV) amplitudes, a solution to the Ward identity may be written as,
\beq\label{eq:Rndef}
A_n^{\textrm{MHV}} = A_n^{\textrm{BDS}} \times \exp(R_n),
\eeq
where $A_n^{\textrm{BDS}}$ is an all-loop, all-multiplicity ansatz proposed by Bern, Dixon, and Smirnov~\cite{Bern2005iz}, and $R_n$ is a dual-conformally invariant function referred to as the {\it remainder function}~\cite{Bern2008ap,Drummond2008aq}.

Dual conformal invariance provides a strong constraint on the form of $R_n$. For example,
it is impossible to construct a non-trivial dual-conformally invariant function with
fewer than six external momenta. As a result, $R_4=R_5=0$, and, consequently,
the four- and five-point scattering amplitudes are equal to the BDS ansatz.
%, whose only ambiguity is a set of momentum-independent constants.
At six points, there are three independent invariant cross ratios built from distances $x_{ij}^2$ 
in the dual space,
\bea\label{eq:cross_ratio_def}
u_1 = \frac{x_{13}^2 x_{46}^2}{x_{14}^2 x_{36}^2}
= \frac{ s_{12} s_{45} }{ s_{123} s_{345} }\,, \qquad
u_2 = \frac{x_{24}^2 x_{15}^2}{x_{25}^2 x_{14}^2}
= \frac{ s_{23} s_{56} }{ s_{234} s_{456} }\,, \qquad
u_3 = \frac{x_{35}^2 x_{26}^2}{x_{36}^2 x_{25}^2}
= \frac{ s_{34} s_{61} }{ s_{345} s_{561} }\, .
\eea
Dual conformal invariance restricts $R_6$ to be a function of
these variables only, i.e. $R_6=R_6(u_1,u_2,u_3)$. This function is not arbitrary since, among other conditions, it must be totally symmetric under permutations of the $u_i$ and vanish in the collinear limit~\cite{Bern2008ap}.

In the absence of an explicit computation, it remained a possibility that $R_6=0$, despite the fact that all known symmetries allow for a non-zero function $R_6(u_1,u_2,u_3)$. However, a series of calculations have since been performed and they showed definitively that $R_6\neq 0$.
The first evidence of a non-vanishing $R_6$ came from an analysis of the multi-Regge
limits of $2\to 4$ gluon scattering amplitudes at two loops~\cite{Bartels2008ce}. Numerical evidence was soon found at specific kinematic points~\cite{Bern2008ap,Drummond2008aq}, and an explicit calculation for general kinematics followed shortly thereafter~\cite{DelDuca2009au,DelDuca2010zg}.
Interestingly, the two-loop calculation for general kinematics was actually performed in a quasi-multi-Regge limit; the full kinematic dependence could then be inferred because this type of Regge limit does not modify the analytic dependence of the remainder function on the $u_i$.

Even beyond the two-loop remainder function, the limit of multi-Regge kinematics (MRK) has received considerable attention in the context of $\cN=4$ super-Yang Mills theory~\cite{Bartels2008ce,Bartels2008sc,Schabinger2009bb,Lipatov2010qg,Bartels2010ej,%
Lipatov2010ad,Bartels2010tx,Dixon2011pw,Fadin2011we,Prygarin2011gd,%
Bartels2011ge,Lipatov2012gk,Dixon2012yy,Bartels2012gq}. One reason for this is that multi-leg scattering amplitudes become considerably simpler in MRK while still maintaining a non-trivial analytic structure.  Taking the multi-Regge limit at six points, for example, essentially reduces the amplitude to a function of just two variables, $w$ and $\ws$, which are complex conjugates of each other. This latter point has proved particularly important in describing the relevant function space in this limit. In fact, it has been argued recently~\cite{Dixon2012yy} that the function space is spanned by the set of single-valued harmonic polylogarithms (SVHPLs) introduced by Brown~\cite{BrownSVHPLs}. These functions will play a prominent role in the remainder of this article.

The MRK limit of $2\to 4$ scattering is characterized by the condition that the outgoing particles are widely separated in rapidity while having comparable transverse momenta. In terms of the cross ratios $u_i$, the limit is approached by sending one of the $u_i$, say $u_1$, to unity, while letting the other two cross ratios vanish at the same rate that $u_1\to 1$, i.e. $u_2 = x(1-u_1)$ and $u_3=y(1-u_1)$ for two fixed variables $x$ and $y$. Actually, this prescription produces the Euclidean version of the MRK limit in which the six-point remainder function vanishes~\cite{Brower2008nm,Brower2008ia,DelDuca2008jg}. To reach the Minkowski version, which is relevant for $2\to 4$ scattering, $u_1$ must be analytically continued around the origin, $u_1\to e^{-2\pi i}|u_1|$, before taking the limit. The remainder function may then be expanded around $u_1 = 1$ and the coefficients of this expansion are functions of only two variables, $x$ and $y$. The variables $w$ and $\ws$ mentioned previously are related to $x$ and $y$ by~\cite{Lipatov2010ad,Bartels2010tx},
\beq\label{eq:xyw_intro}
x\equiv {1\over (1+w)(1+\ws)}, \qquad y\equiv {w\,\ws\over (1+w)(1+\ws)}\,.
\eeq

Neglecting terms that vanish like powers of $1-u_1$, the expansion of the remainder function may be written as\footnote{We follow the conventions of ref.~\cite{Dixon2011pw}.},
\beq\label{eq:R6_MRK_intro}
R_6^{\textrm{MHV}}|_{\textrm{MRK}}
\,=\, 2\pi i\,\sum_{\ell=2}^\infty\sum_{n=0}^{\ell-1}a^\ell\,\log^n(1-u_1)\,
\left[g_n^{(\ell)}(w,\ws) + 2\pi i\,h_n^{(\ell)}(w,\ws)\right]\,,
\eeq
where the coupling constant for planar $\cN=4$ super-Yang-Mills theory
is $a = g^2 N_c/(8\pi^2)$. This expansion is organized hierarchically into
the leading-logarithmic approximation (LLA) with $n=\ell-1$, 
the next-to-leading-logarithmic approximation (NLLA) with $n=\ell-2$,
and in general the N$^k$LL terms with $n=\ell-k-1$. In this article, we study the
 leading-logarithmic approximation, for which we may rewrite eq.~\eqref{eq:R6_MRK_intro} as,
\beq\label{eq:R6_LLA_intro}
R_6^{\textrm{MHV}}|_{\textrm{LLA}}
\,=\, \frac{2\pi i}{\log(1-u_1)}\,\sum_{\ell=2}^\infty \eta^\ell\,g_{\ell-1}^{(\ell)}(w,\ws) \, ,
\eeq
where we have identified $\eta= a \log(1-u_1)$ as the relevant expansion parameter. In LLA, the real part of $R_6$ vanishes, so $h_{\ell-1}^{\ell}(w,\ws)$ is absent in eq.~\eqref{eq:R6_LLA_intro}. Expressions for $g_{\ell-1}^{(\ell)}(w,\ws)$ have been given in the literature for two, three~\cite{Lipatov2010ad}, and recently up to ten~\cite{Dixon2012yy} loops.

An all-orders integral-sum representation for $R_6^{\textrm{MHV}}|_{\textrm{LLA}}$ was presented in ref.~\cite{Lipatov2010ad} and was generalized to the NMHV helicity configuration in ref.~\cite{Lipatov2012gk}. (The MHV case was extended to NLLA in ref.~\cite{Fadin2011we}.) The formula may be understood as an inverse Fourier-Mellin transform from a space of moments labeled by $(\nu,n)$ to the space of kinematic variables $(w,\ws)$. In the moment space, $R_6|_{\textrm{LLA}}(\nu,n)$ assumes a simple factorized form and may be written succinctly to all loop orders in terms of polygamma functions. This structure is obscured in $(w,\ws)$ space, as the inverse Fourier-Mellin transform generates complicated combinations of polylogarithmic functions. Nevertheless, these complicated expressions should bear the mark of their simple ancestry. In this article, we expose this inherited structure by presenting an explicit all-orders formula for $R_6|_{\textrm{LLA}}$ directly in $(w,\ws)$ space.

We do not present a proof of this formula, but we do test its validity using several non-trivial consistency checks. For example, our result agrees with the integral formula mentioned above through at least 14 loops. In ref.~\cite{Lipatov2012gk}, Lipatov, Prygarin, and Schnitzer give a simple differential equation linking the MHV and NMHV helicity configurations,
\beq\label{eq:MHV_NMHV_diffeq_intro}
\ws \frac{\partial}{\partial \ws} R_6^{\textrm{MHV}}|_{\textrm{LLA}}  = w \frac{\partial}{\partial w} R_6^{\textrm{NMHV}}|_{\textrm{LLA}}\, ,
\eeq
which is also obeyed our formula. In the near-collinear limit, we find agreement with the all-orders double-leading-logarithmic approximation of Bartels, Lipatov, and Prygarin~\cite{Bartels2011xy}. 

This article is organized as follows. In Section~\ref{sec:MRK}, we review the aspects of multi-Regge kinematics relevant to six-particle scattering and recall the integral formulas for $R_6|_{\textrm{LLA}}$ in the MHV and NMHV helicity configurations. The construction and properties of single-valued harmonic polylogarithms are reviewed in Section~\ref{sec:SVHPLs}. An all-orders expression for $R_6|_{\textrm{LLA}}$ is presented in terms of these functions in Section~\ref{sec:R6LLA}. After verifying several consistency conditions of this formula, we examine its near-collinear limit in Section~\ref{sec:coll}. Section~\ref{sec:conclusion} offers some concluding remarks and prospects for future work.

%%%%%%%%%%%%%%%%%%%%%%%%%%%%%%%%%%%%
%%%%%%%%%%%%%%%%%%%%%%%%%%%%%%%%%%%%
\section{The six-point remainder function in multi-Regge kinematics}
\label{sec:MRK}
We consider the six-gluon scattering process $g_3g_6\to g_1g_5g_4g_2$ where the momenta are taken to be outgoing and the gluons are labeled cyclically in the clockwise direction. The limit of multi-Regge kinematics is defined by the condition that the produced gluons are strongly ordered in rapidity while having comparable transverse momenta,
\beq
y_1 \gg y_5 \gg y_4 \gg y_2\,,\quad\quad |p_{1\perp}|\simeq|p_{5\perp}|\simeq|p_{4\perp}|\simeq|p_{2\perp}|\, .
\eeq
In the Euclidean region, this limit is equivalent to the hierarchy of scales,
\beq
s_{12} \gg s_{345},\,s_{456} \gg s_{34},\, s_{45}\,,s_{56} 
      \gg s_{23},\, s_{61},\,s_{234},
\eeq
which leads to the limiting behavior of the cross ratios~\eqref{eq:cross_ratio_def},
\beq
1-u_1,\, u_2,\, u_3 \sim 0\,,
\eeq
subject to the constraint that the following ratios are held fixed,
\beq\label{eq:xy_def}
x\equiv {u_2\over1-u_1} = \cO(1) {\rm~~and~~} 
y\equiv {u_3\over1-u_1} = \cO(1)\,.
\eeq
Unitarity restricts the branch cuts of physical quantities like the remainder function $R_6(u_1,u_2,u_3)$ 
to appear in physical channels. In terms of the cross ratios $u_i$, this requirement implies that all branch points occur
when a cross ratio vanishes or approaches infinity. If we re-express the two real variables $x$ and $y$ by a single complex variable $w$,
\beq\label{eq:xy_to_wws}
x\equiv {1\over (1+w)(1+\ws)} {\rm~~and~~} y\equiv {w\,\ws\over (1+w)(1+\ws)}\,,
\eeq
then the equivalent statement in MRK is that any function of $(w,\ws)$ must be \emph{single-valued} in the complex $w$ plane.

In the Euclidean region, the remainder function actually vanishes in the multi-Regge limit. To obtain a non-vanishing result, we must consider a physical region in which one of the cross ratios acquires a phase~\cite{Bartels2008ce}. One such region corresponds to the $2\to4$ scattering process described above. It can be reached by flipping the signs of $s_{12}$ and $s_{45}$, or, in terms of the cross ratios, by rotating $u_1$ around the origin,
\beq\label{eq:MRK_anal_cont}
u_1 \to e^{-2\pi i}\,|u_1|\, .
\eeq
In the course of this analytic continuation, we pick up the discontinuity across a Mandelstam cut~\cite{Bartels2008ce,Lipatov2010qg}. The six-point remainder function can then be expanded in the form given in eq.~\eqref{eq:R6_MRK_intro},
\beq\label{eq:R6_MRK}
R_6^{\textrm{MHV}}|_{\textrm{MRK}} = 2\pi i\,\sum_{\ell=2}^\infty\sum_{n=0}^{\ell-1}a^\ell\,\log^n(1-u_1)\,\left[g_n^{(\ell)}(w,\ws) + 2\pi i\,h_n^{(\ell)}(w,\ws)\right]\,.
\eeq
The large logarithms $\log(1-u_1)$ organize this expansion into the leading-logarithmic approximation (LLA) with $n=\ell-1$, the next-to-leading-logarithmic approximation (NLLA) with $n=\ell-2$, and in general the the N$^k$LL terms with $n=\ell-k-1$.

 In refs.~\cite{Lipatov2010ad, Fadin2011we} an all-loop integral formula for 
$R_6^{\textrm{MHV}}|_{\textrm{MRK}}$ was presented for LLA and NLLA\footnote{There is a difference
in conventions regarding the definition of the remainder function. What we
call $R$ is called $\log(R)$ in refs.~\cite{Lipatov2010ad, Fadin2011we}.
Apart from the zeroth order term, this distinction has no effect on LLA terms. The first place it makes a difference
is at four loops in NLLA, in the real part.},
\beq\label{eq:MHV_MRK}
e^{R+i\pi\delta}|_{\textrm{MRK}} = \cos\pi\omega_{ab} 
+ i \, {a\over 2} \sum_{n=-\infty}^\infty
(-1)^n\,\left({w\over \ws}\right)^{{n\over 2}}\int_{-\infty}^{+\infty}
{d\nu\over \nu^2+{n^2\over 4}}\,|w|^{2i\nu}\,\Phi_{\textrm{Reg}}(\nu,n)
\,\left(-{1\over \sqrt{u_2\,u_3}}\right)^{\omega(\nu,n)} .
\eeq
Here, $\omega(\nu,n)$ is the BFKL eigenvalue and $\Phi_{\textrm{Reg}}(\nu,n)$ is the regularized impact factor. They may be expanded perturbatively,
\beq\bsp\label{eq:omega_phi}
\omega(\nu,n) &\,= 
- a \left(E_{\nu,n} + a\,E_{\nu,n}^{(1)}+ a^2\,E_{\nu,n}^{(2)}+\cO(a^3)\right)\,,\\
\Phi_{\textrm{Reg}}(\nu,n)&\, = 1 + a \, \Phi_{\textrm{Reg}}^{(1)}(\nu,n)
 + a^2 \, \Phi_{\textrm{Reg}}^{(2)}(\nu,n)
 + a^3 \, \Phi_{\textrm{Reg}}^{(3)}(\nu,n)+\cO(a^4)\,.
\esp\eeq
The leading-order eigenvalue, $E_{\nu,n}$, was given in ref.~\cite{Bartels2008sc} and may be written in terms of the digamma function $\psi(z) = {d\over dz}\log\Gamma(z)$,
\beq
E_{\nu,n} = -{1\over2}\,{|n|\over \nu^2+{n^2\over 4}}
+\psi\left(1+i\nu+{|n|\over2}\right) +\psi\left(1-i\nu+{|n|\over2}\right) - 2\psi(1) \, .
\eeq
In this article, we will only need the leading-order terms, but, remarkably, the higher-order corrections listed  in~\eqref{eq:omega_phi} may also be expressed in terms of the $\psi$ function and its derivatives~\cite{Fadin2011we,Dixon2012yy}.

Returning to~\eqref{eq:MHV_MRK}, the remaining functions are,
\beq\bsp
\omega_{ab}& \; =\; {1\over 8}\,\gamma_K(a)\,\log{u_3\over u_2}
 = {1\over 8}\,\gamma_K(a)\,\log|w|^2\,, \\
\delta &\;=\; {1\over 8}\,\gamma_K(a)\,\log{(xy)}
 = {1\over 8}\,\gamma_K(a)\,\log{|w|^2\over |1+w|^4}\,,
\esp\eeq
and the cusp anomalous dimension, which is known to all orders in
perturbation theory~\cite{BES},
\be\label{eq:gamma_cusp}
\gamma_K(a) \, = \, \sum_{\ell=1}^\infty \gamma_K^{(\ell)} a^\ell
           \,  = \, 4\,a - 4\,\zeta_2 \, a^2 + 22 \, \zeta_4 \, a^3 
             - ( \textstyle{\frac{219}{2}} \, \zeta_6 + 4 \, \zeta_3^2 ) \, a^4
             + \cdots \,. 
\ee
In addition, there is an ambiguity regarding the Riemann sheet of the exponential
factor on the right-hand side of~\eqref{eq:MHV_MRK}. We resolve this ambiguity
with the identification,
\beq\label{eq:ipi}
\left(-{1\over \sqrt{u_2\,u_3}}\right)^{\omega(\nu,n)}
\to e^{-i\pi\omega(\nu,n)}\,\left( {1\over 1-u_1}\,{|1+w|^2\over |w|}\right)^{\omega(\nu,n)}\, .
\eeq

The $i\pi$ factor in the right-hand side of eq.~\eqref{eq:ipi} generates the real parts $h_n^{(\ell)}$ in eq.~\eqref{eq:R6_MRK}. For example, at LLA and NLLA, the following relations~\cite{Dixon2012yy} are satisfied\footnote{Note that the sum over $k$ in the formula for $h_{\ell-2}^{(\ell)}$ would not have been present if we had used the convention for $R$ in 
refs.~\cite{Lipatov2010ad, Fadin2011we}.},
\beq\bsp\label{eq:htog}
h_{\ell-1}^{(\ell)}(w,\ws) &\,= 0\,, \\
h_{\ell-2}^{(\ell)}(w,\ws) &\,= 
{\ell-1\over2} \, g_{\ell-1}^{(\ell)}(w,\ws)
+ {1\over 16} \, \gamma_K^{(1)} \, g_{\ell-2}^{(\ell-1)}(w,\ws)
\,\log{|1+w|^4 \over|w|^2} \\
 &\quad - \frac{1}{2} \sum_{k=2}^{\ell-2} g_{k-1}^{(k)} g_{\ell-k-1}^{(\ell-k)} \,, 
\qquad \ell > 2,
\esp\eeq
where $\gamma_K^{(1)} = 4$ from eq.~(\ref{eq:gamma_cusp}). Making use of eq.~\eqref{eq:R6_LLA_intro}, we present an alternate form of these identities which will be useful later,
\beq\bsp\label{eq:ReNLLA}
\textrm{Re}\left(R_6^{\textrm{MHV}}|_{\textrm{NLLA}}\right) & \;=\; \frac{2\pi i}{\log(1-u_1)}\bigg(\frac{1}{2}\,\eta^2\,\frac{\partial}{\partial \eta}\,\frac{1}{\eta} + \frac{\gamma_K^{(1)}}{16}\,\eta\,\log\frac{|1+w|^4}{|w|^2}\, \bigg)\,R_6^{\textrm{MHV}}|_{\textrm{LLA}} \\
&\quad   + \frac{2\pi^2}{\log^2(1-u_1)}\eta^2 g^{(2)}_1(w,\ws) - \frac{1}{2}\left(R_6^{\textrm{MHV}}|_{\textrm{LLA}}\right)^2\,.
\esp\eeq
The term proportional to $g^{(2)}_1(w,\ws)$ addresses the special case of $\ell=2$ in eq.~\eqref{eq:htog}.

In what follows, we will focus on the leading-logarithmic approximation of~\eqref{eq:MHV_MRK}, which takes the form,
\beq\label{eq:MHV_MRK_LLA}
R_6^{\textrm{MHV}}|_{\textrm{LLA}}  =  i{a\over 2}
\sum_{n=-\infty}^\infty(-1)^n\,
\int_{-\infty}^{+\infty}
{d\nu \, w^{i\nu+n/2} \, \ws^{i\nu-n/2} \over (i\nu+{n\over 2})(-i\nu+{n\over 2})}
\,\bigg[(1-u_1)^{a\,E_{\nu,n}}-1\bigg] \,.
\eeq
The $\nu$-integral may be evaluated by closing the contour and summing residues\footnote{For the special case of $n=0$, our prescription is to take half the residue at $\nu=0$.}. To perform the resulting double sums, one may apply the summation algorithms of ref.~\cite{Moch2001zr}, although this approach is computationally challenging for high loop orders. Alternatively, an ansatz for the result may be expanded around $|w|=0$ and matched term-by-term to the truncated double sum. The latter method requires knowledge of the complete set of functions that might arise in this context. In ref.~\cite{Dixon2012yy}, it was argued that the single-valued harmonic polylogarithms (SVHPLs) completely characterize this function space, and, using these functions, eq.~\eqref{eq:MHV_MRK_LLA} was evaluated through ten loops. %The SVHPLs will feature prominently throughout this article, and we review their construction and properties in the next section.

So far we have only discussed the MHV helicity configuration. We now turn to the only other independent helicity configuration at six points, the NMHV configuration. In MRK, the MHV and NMHV tree amplitudes are equal~\cite{DelDuca1995zy,Lipatov2012gk}. It is natural, therefore, to define an NMHV remainder function, analogous to eq.~\eqref{eq:Rndef},
\be\label{eq:RdefNMHV}
A_6^{\textrm{NMHV}}|_{\textrm{MRK}} = A_6^{\textrm{BDS}} \times \exp(R_{\textrm{NMHV}}) \,.
\ee
In ref.~\cite{Lipatov2012gk}, it was argued that the effect of changing the helicity of one of the positive-helicity gluons\footnote{Up to power-suppressed terms, helicity must be conserved along high-energy lines, so 
the helicity flip must occur on one of the lower-energy legs, 4 or 5.} was equivalent to changing the impact factor
for that gluon by means of the following replacement,
\beq
{1\over-i\nu+{n\over2}}\to-{1\over i\nu+{n\over2}}\, .
\eeq
Referring to eq.~\eqref{eq:MHV_MRK_LLA}, this replacement leads to an integral formula for $R_6^{\textrm{NMHV}}|_{\textrm{LLA}} $,
\beq\label{eq:NMHV_MRK}
R_6^{\textrm{NMHV}}|_{\textrm{LLA}}  = -{ia\over2}\,\sum_{n=-\infty}^\infty(-1)^n
\,\int_{-\infty}^{+\infty}
{d\nu  \, w^{i\nu+n/2} \, \ws^{i\nu-n/2} \over (i\nu+{n\over 2})^2}\,\bigg[(1-u_1)^{a\,E_{\nu,n}}-1\bigg]\,.
\eeq
Following refs.~\cite{Lipatov2012gk} and~\cite{Dixon2012yy}, we can extract a simple rational prefactor and write eq.~\eqref{eq:NMHV_MRK} in a manifestly inversion-symmetric form,
\beq\label{eq:R6NMHV_LLA_f}
R_6^{\textrm{NMHV}}|_{\textrm{LLA}} = \frac{2\pi i}{\log(1-u_1)}\,\sum_{\ell=2}^\infty \frac{\eta^\ell}{1+\ws}f^{(\ell)}(w,\ws)  \; + \; \bigg\{(w,\ws) \leftrightarrow \left(\frac{1}{w},\frac{1}{\ws}\right)\bigg\}\, ,
\eeq
for some single-valued functions $f^{(\ell)}(w,\ws)$. It is possible to obtain expressions for $f^{(\ell)}(w,\ws)$ directly from eq.~\eqref{eq:NMHV_MRK} by means of the truncated series approach outlined above, for example. A simpler method is to make use of the following differential equation, which may be deduced by comparing the two expressions~\eqref{eq:MHV_MRK_LLA} and~\eqref{eq:NMHV_MRK},
\beq\label{eq:MHV_NMHV_diffeq}
\ws \frac{\partial}{\partial \ws} R_6^{\textrm{MHV}}|_{\textrm{LLA}}  = w \frac{\partial}{\partial w} R_6^{\textrm{NMHV}}|_{\textrm{LLA}}\, .
\eeq
In principle, solving this equation requires the difficult step of fixing the constants of integration in such a way that single-valuedness is preserved.  As discussed in ref.~\cite{Dixon2012yy}, this step becomes trivial when working in the space of SVHPLs, which are the subject of the next section.

%%%%%%%%%%%%%%%%%%%%%%%%%%%%%%%%%%%%%%%%%%
%%%%%%%%%%%%%%%%%%%%%%%%%%%%%%%%%%%%%%%%%%
\section{Review of single-valued harmonic polylogarithms}
\label{sec:SVHPLs}
Harmonic polylogarithms (HPLs)~\cite{Remiddi1999ew} are a class of generalized polylogarithmic functions that finds frequent application in multi-loop calculations. The HPLs are functions of a single complex variable, $z$, which will be related to the kinematic variable $w$ by $z=-w$. We will continue to use $z$ throughout this section in order to make contact with the existing mathematical literature. In general, the HPLs have branch cuts that originate at $z=-1$, $z=0$, or $z=1$. In the present application, we will consider the restricted class of HPLs\footnote{In the mathematical literature, these functions are sometimes referred to as \emph{multiple polylogarithms in one variable}. With a small abuse of notation, we will continue to use the term ``HPL" to refer to this restricted set of functions.} whose branch points are either $z=0$ or $z=1$. To construct them, consider the set $X^*$ of all words $w$ formed from the letters $x_0$ and $x_1$, together with $e$, the empty word\footnote{Context should distinguish the word $w$ from the kinematic variable with the same name.}. Then, for each $w\in X^*$, define a function $H_w(z)$ which obeys the differential equations,
\begin{equation}\label{eq:HPLdef_1}
\frac{\partial}{\partial z}H_{x_0w}(z) = \frac{H_{w}(z)}{z} \quad\quad\text{and}\quad\quad\frac{\partial}{\partial z}H_{x_1w}(z) = \frac{H_{w}(z)}{1-z}\,,
\end{equation}
subject to the following conditions,
\begin{equation}
H_{e}(z)=1,\quad\quad H_{x_0^n}(z)=\frac{1}{n!}\log^nz,\quad\quad\text{and}\quad\quad \lim_{z\rightarrow 0}H_{w\neq x_0^n}(z)=0\,.
\end{equation}
There is a unique family of solutions to these equations, and it defines the HPLs. For $w\neq x_0^n$, they can be written as iterated integrals,
\begin{equation}\label{eq:HPL_integral}
H_{x_0 w}(z) = \int_0^z dz'\; \frac{H_w(z')}{z'} \quad\quad \text{and}\quad\quad H_{x_1w}= \int_0^z dz'\; \frac{H_w(z')}{1-z'}\,.
\end{equation}
The structure of the iterated integrals endows the HPLs with an important property: they form a \emph{shuffle algebra}. The shuffle relations can be written as,
\begin{equation}
\label{eq:HPL_shuffle}
H_{w_1}(z)\,H_{w_2}(z) = \sum_{{w}\in{w_1}\ssha {w_2}}H_{w}(z)\,,
\end{equation}
where ${w_1}\sha{w_2}$ is the set of mergers of the sequences $w_1$ and $w_2$ that preserve their relative ordering. The shuffle algebra may be used to remove all zeros from the right of an index vector in favor of some explicit logarithms. For example, it is easy to obtain the following formula for HPLs with a single $x_1$,
\beq\label{eq:shuff_x1}
H_{x_0^n x_1 x_0^m} = \sum_{j=0}^{m}\frac{(-1)^{j}}{(m-j)!}\binom{n+j}{j}H_{x_0}^{m-j} H_{x_0^{n+j}x_1}\,.
\eeq
After removing all right-most zeros, the Taylor expansions around $z=0$ are particularly simple and involve only a
special class of harmonic numbers~\cite{Remiddi1999ew},
\beq\label{eq:HPL_series}
H_{m_1,\ldots,m_k}(z) = \sum_{l=1}^\infty{z^l\over l^{m_1}}Z_{m_2,\ldots,m_k}(l-1)\,,
\qquad m_i>0\,,
\eeq
where $Z_{m_1,\ldots,m_k}(n)$ are Euler-Zagier
sums~\cite{Euler_sum,Zagier_sum}, defined recursively by
\beq\label{eq:Euler-Zagier}
Z(n) =1 {\rm~~and~~}
Z_{m_1,\ldots,m_k}(n) = \sum_{l=1}^n{1\over l^{m_1}}Z_{m_2,\ldots,m_k}(l-1)\,.
\eeq
Note that the indexing of the weight vectors $m_1,\ldots,m_k$ in
eqs.~(\ref{eq:HPL_series}) and (\ref{eq:Euler-Zagier}) is in the
collapsed notation in which a subscript $m$ denotes $m-1$ zeros followed by a single $1$.

The HPLs are multi-valued functions; nevertheless, it is possible to build specific combinations such that the branch cuts cancel and the result is single-valued. An algorithm that explicitly constructs these combinations was presented in ref.~\cite{BrownSVHPLs} and reviewed in ref.~\cite{Dixon2012yy}. Here we provide a very brief description.

The SVHPLs $ \cL_w(z)$ are generated by the series,
\beq\label{eq:Lgen}
\cL(z)=L_X(z)\tilde{L}_Y(\bar{z}) \equiv \sum_{w\in X^*} \cL_w(z)w\, ,
\eeq
where,
\beq\label{eq:LXY0}
L_X(z)\,=\,\sum_{w\in X^*}H_w(z)w \,, \qquad
\tilde{L}_Y(\bar{z})\,=\,\sum_{w\in Y^*}H_{\phi(w)}(\bar{z})\tilde{w} \,.
\eeq
Here $^ \sim: X^*\rightarrow X^*$ is the operation that reverses words, $\phi:Y^*\rightarrow X^*$ is the map that renames $y$ to $x$, and $Y^*$ is the set of words in $\{y_0,y_1\}$, which are defined by the relations,
\beq\bsp
\label{eq:y_alph}
y_0&\,=\,x_0\\
\tilde{Z}(y_0,y_1)y_1\tilde{Z}(y_0,y_1)^{-1} &\,=\, Z(x_0,x_1)^{-1}x_1 Z(x_0,x_1),
\esp\eeq
where $Z(x_0,x_1)$ is a generating function of multiple zeta values,
\begin{equation}
Z(x_0,x_1) = \sum_{w\in X^*} \zeta(w)w.
\end{equation}
The $\zeta(w)$ are regularized by the shuffle algebra and obey $\zeta(w\neq x_1)=H_{w}(1)$ and $\zeta(x_1)=0$.

Alternatively, one may formally define these functions as solutions to simple differential equations, i.e. the $\cL_w(z)$ are the unique single-valued linear combinations of functions $H_{w_1}(z)H_{w_2}(\bar{z})$ that obey the differential equations~\cite{BrownSVHPLs},
\begin{equation}\label{eq:Lzdiffeq}
\frac{\partial}{\partial z}\cL_{x_0w}(z) = \frac{\cL_{w}(z)}{z} \quad\quad\text{and}\quad\quad\frac{\partial}{\partial z}\cL_{x_1w}(z) = \frac{\cL_{w}(z)}{1-z}\,,
\end{equation}
subject to the conditions,
\begin{equation}
\cL_{e}(z)=1\,,\quad\quad \cL_{x_0^n}(z)=\frac{1}{n!}\log^n|z|^2\quad\quad\text{and}\quad\quad \lim_{z\rightarrow 0}\cL_{w\neq x_0^n}(z)=0\,.
\end{equation}
The SVHPLs also obey differential equations in $\bar{z}$. Both sets of equations are represented nicely in terms of the generating function~\eqref{eq:Lgen},
\beq\label{eq:dLzz}
\frac{\partial}{\partial z} \cL(z) = \left(\frac{x_0}{z}+\frac{x_1}{1-z}\right)\cL(z) \quad\quad \textrm{and}\quad\quad\frac{\partial}{\partial \bar{z}} \cL(z) = \cL(z)\left(\frac{y_0}{\bar{z}}+\frac{y_1}{1-\bar{z}}\right).
\eeq
%%%%%%%%%%%%%%%%%%%%%%%%%%%%%%%%%%%%%%%%%%%%%%%%%%%%%%%%%%%%%%%%%%%%
%%%%%%%%%%%%%%%%%%%%%%%%%%%%%%%%%%%%%%%%%%%%%%%%%%%%%%%%%%%%%%%%%%%%
\section{Six-point remainder function in MRK and LLA}
\label{sec:R6LLA}
The SVHPLs introduced in the previous section provide a convenient basis of functions to describe the six-point remainder function in MRK. In ref.~\cite{Dixon2012yy}, these functions were used to express the result through ten loops in LLA and through nine loops in NLLA. Here we use the SVHPLs to present a formula in LLA to all loop orders.
\subsection{The all-orders formula}
Recall from the previous section that we defined $X^*$ to be the set of all words $w$ in the letters $x_0$ and $x_1$ together with the empty word $e$. Let $\mathbb{C}\langle X\rangle$ be the complex vector space generated by $X^*$ and let $\mathbb{C}\langle \cL\rangle$ be the complex vector space spanned by the SVHPLs, $\cL_w$ with $w\in X^*$. Denote by $\mathbb{C}\langle X\rangle[[\eta]]$ and $\mathbb{C}\langle \cL\rangle[[\eta]]$ the rings of formal power series in the variable $\eta=a \log(1-u_1)$ with coefficients in $\mathbb{C}\langle X\rangle$ and $\mathbb{C}\langle \cL\rangle$, respectively. There is a natural map, $\rho$, which sends words to the corresponding SVHPLs,
\beq\bsp
\rho: \mathbb{C}\langle X\rangle[[\eta]] &\,\rightarrow \,\mathbb{C}\langle \cL\rangle[[\eta]]\\
w &\,\mapsto \, \cL_w\, .
\esp\eeq
Using these ingredients, we propose the following formulas for the MHV and NMHV remainder functions in MRK and LLA,
\begin{eqnarray}
\label{R6eqnMHV}
R_6^{\textrm{MHV}}|_{\textrm{LLA}} &=& \frac{2\pi i}{\log(1-u_1)}\,\rho\Big(\mathcal{X}\cZ^{\textrm{MHV}}- \frac{1}{2}\,x_1 \eta\Big)\, ,\\
\label{R6eqnNMHV}
R_6^{\textrm{NMHV}}|_{\textrm{LLA}}&=& \frac{2\pi i}{\log(1-u_1)}\,\frac{1}{1+\ws}\,\rho\Big(x_0\mathcal{X} \cZ^{\textrm{NMHV}}\Big) \; + \; \bigg\{(w,\ws) \leftrightarrow \left(\frac{1}{w},\frac{1}{\ws}\right)\bigg\}\, ,
\end{eqnarray}
where the formal power series $\mathcal{X}, \cZ^{\textrm{(N)MHV}} \in \mathbb{C}\langle X\rangle[[\eta]]$ are,
\beq\bsp\label{eq:XZZ}
%\,=\,\sum_{k=0}^{\infty}\left(\sum_{n=0}^{k}\frac{x_0^{k-n}}{2^{k-n}\,(k-n)!}\sum_{\alpha\in Q(n)}\prod_j\frac{x_1x_0^{\alpha_j-1}}{\alpha_j!}\right)y^k
\mathcal{X} &\,=\,e^{\frac{1}{2}x_0 \eta}\left[1-x_1 \left(\frac{e^{x_0 \eta}-1}{x_0}\right)\right]^{-1}\,,\\
\cZ^{\textrm{MHV}}&\,=\,\frac{1}{2}\sum_{k=1}^{\infty}\left(x_1\, \sum_{n=0}^{k-1}(-1)^n x_0^{k-n-1}\sum_{m=0}^{n} \, \frac{2^{2m-k+1}}{(k-m-1)!} \,\mathfrak{Z}(n,m)\right)\eta^k\, , \\
\cZ^{\textrm{NMHV}}&\,=\,\frac{1}{2}\sum_{k=2}^{\infty}\left(x_1\, \sum_{n=0}^{k-2}(-1)^n x_0^{k-n-2}\sum_{m=0}^{n} \, \frac{2^{2m-k+1}}{(k-m-1)!} \,\mathfrak{Z}(n,m)\right)\eta^k \, .
\esp\eeq
Here, the $\mathfrak{Z}(n,m)$ are particular combinations of $\zeta$ values of uniform weight $n$. They are related to partial Bell polynomials, and are generated by the series,
\beq
 \exp\left[y \sum_{k=1}^{\infty} \zeta_{2k+1} x^{2k+1}\right] \equiv \sum_{n=0}^\infty \sum_{m=0}^\infty \mathfrak{Z}(n,m)\,x^n y^m\, .
\eeq
An explicit formula is,
\beq
\mathfrak{Z}(n,m) \;\;= \sum_{\beta \in P(n,m)} \prod_i \frac{(\zeta_{2i+1})^{\beta_i}}{\beta_i!}\, ,
\eeq
where $P(n,m)$ is the set of $n$-tuples of non-negative integers that sum to $m$, such that the product of $\zeta$ values has weight $n$,
\beq
P(n,m) \,=\, \left\{\{\beta_1,\cdots,\beta_n\}  \,\Big|\,\beta_i \in \mathbb{N}_0,\; \sum_{i=1}^{n} \beta_i = m,\; \sum_{i=1}^{n} (2i+1)\beta_i = n \right\}\, .
\eeq
Similarly, an expression for the $k$th term of $\mathcal{X}$ can be given as,
\beq
\mathcal{X} = \sum_{k=0}^{\infty}\left(\sum_{n=0}^{k}\frac{x_0^{k-n}}{2^{k-n}\,(k-n)!}\sum_{\alpha\in Q(n)}\prod_j\frac{x_1x_0^{\alpha_j-1}}{\alpha_j!}\right)\eta^k\,,
\eeq
where $Q(n)$ is the set of integer compositions of $n$,
\beq
Q(n) \,=\, \left\{\{\alpha_1, \alpha_2, \cdots,\alpha_m\}  \,\Big|\,\alpha_i \in \mathbb{Z}^{+},\; \sum_{i=1}^{m} \alpha_i = n\right\}\, .
\eeq

Excluding the one-loop term in eq.~\eqref{R6eqnMHV}, the arguments of the $\rho$ functions factorize into the product of a $\zeta$-free function, $\mathcal{X}$, and a $\zeta$-containing function, $\cZ^{\textrm{(N)MHV}}$. The $\zeta$-free function is simpler and its first few terms read,
\beq\bsp\label{eq:Xexp}
\mathcal{X} &\;=\; 1+\left(\frac{1}{2}\,x_0 + x_1\right)\eta + \left(\frac{1}{8}\,x_0^2+\frac{1}{2}\,x_0x_1+\frac{1}{2}\,x_1x_0+x_1^2\right)\eta^2 \\
&\quad +\left(\frac{1}{48}\,x_0^3+\frac{1}{8}\,x_0^2x_1+\frac{1}{4}\,x_0x_1x_0+\frac{1}{2}\,x_0x_1^2+\frac{1}{6}\,x_1x_0^2+\frac{1}{2}\,x_1x_0x_1+\frac{1}{2}\,x_1^2x_0+x_1^3\right)\eta^3 +  \cdots\, .
\esp\eeq
The $\zeta$-containing functions are slightly more complicated. Their first few terms are,
\beq\bsp\label{eq:Zexp}
\cZ^{\textrm{MHV}} &\;=\; \frac{1}{2}\, x_1\,\eta + \frac{1}{4}\,x_1x_0\,\eta^2+\frac{1}{16}\,x_1x_0^2\,\eta^3 +\left(\frac{1}{96}\,x_1x_0^3-\frac{1}{8}\,\zeta_3\,x_1\right)\eta^4 + \cdots \, ,\\
\cZ^{\textrm{NMHV}} &\;=\; \frac{1}{4}\, x_1\,\eta^2 + \frac{1}{16}\,x_1x_0\,\eta^3+\frac{1}{96}\,x_1x_0^2\,\eta^4 +\left(\frac{1}{768}\,x_1x_0^3-\frac{1}{48}\,\zeta_3\,x_1\right)\eta^5 + \cdots \, .
\esp\eeq
Using eqs.~\eqref{eq:Xexp} and~\eqref{eq:Zexp}, one may easily extract $g^{(\ell)}_{\ell-1}$ for $\ell=1,2,3,4$ (cf. eqs.~\eqref{eq:R6_LLA_intro} and~\eqref{R6eqnMHV}). The one loop term vanishes, $g^{(1)}_0=0$, and the other functions read,
\beq\bsp\label{eq:g_low_orders}
g^{(2)}_{1} &\;=\;\frac{1}{4}\,\cL_{0,1}+\frac{1}{4}\,\cL_{1,0}+\frac{1}{2}\,\cL_{1,1}\, , \\
g^{(3)}_{2} &\;=\;\frac{1}{16}\,\cL_{0,0,1}+\frac{1}{8}\,\cL_{0,1,0}+\frac{1}{4}\,\cL_{0,1,1}+\frac{1}{16}\,\cL_{1,0,0}+\frac{1}{4}\,\cL_{1,0,1}+\frac{1}{4}\,\cL_{1,1,0}+\frac{1}{2}\,\cL_{1,1,1}\, , \\
g^{(4)}_{3} &\;=\;\frac{1}{96}\,\cL_{0,0,0,1}+\frac{1}{32}\,\cL_{0,0,1,0}+\frac{1}{16}\,\cL_{0,0,1,1}+\frac{1}{32}\,\cL_{0,1,0,0}+\frac{1}{8}\,\cL_{0,1,0,1}+\frac{1}{8}\,\cL_{0,1,1,0}\\
&\qquad+\frac{1}{4}\,\cL_{0,1,1,1}+\frac{1}{96}\,\cL_{1,0,0,0}+\frac{1}{12}\,\cL_{1,0,0,1}\, +\frac{1}{8}\,\cL_{1,0,1,0}+\frac{1}{4}\,\cL_{1,0,1,1}+\frac{1}{16}\,\cL_{1,1,0,0} \\
&\qquad+\frac{1}{4}\,\cL_{1,1,0,1}+\frac{1}{4}\,\cL_{1,1,1,0}+\frac{1}{2}\,\cL_{1,1,1,1}-\frac{1}{8}\,\zeta_3\,\cL_{1}\, .
\esp\eeq
Similarly, one may extract the first few $f^{(\ell)}$ (cf. eqs.~\eqref{eq:R6NMHV_LLA_f} and~\eqref{R6eqnNMHV}), finding $f^{(1)}=0$ and,
\beq\bsp\label{eq:f_low_orders}
f^{(2)} &\;=\;\frac{1}{4}\,\cL_{0,1}\, , \\
f^{(3)} &\;=\;\frac{1}{8}\,\cL_{0,0,1}+\frac{1}{16}\,\cL_{0,1,0}+\frac{1}{4}\,\cL_{0,1,1}\, , \\
f^{(4)} &\;=\;\frac{1}{32}\,\cL_{0,0,0,1}+\frac{1}{32}\,\cL_{0,0,1,0}+\frac{1}{8}\,\cL_{0,0,1,1}+\frac{1}{96}\,\cL_{0,1,0,0}+\frac{1}{8}\,\cL_{0,1,0,1}+\frac{1}{16}\,\cL_{0,1,1,0}+\frac{1}{4}\,\cL_{0,1,1,1}\, , \\
f^{(5)} &\;=\;\frac{1}{192}\,\cL_{0,0,0,0,1}+\frac{1}{128}\,\cL_{0,0,0,1,0}+\frac{1}{32}\,\cL_{0,0,0,1,1}+\frac{1}{192}\,\cL_{0,0,1,0,0}+\frac{1}{16}\,\cL_{0,0,1,0,1}\\
&\qquad+\frac{1}{32}\,\cL_{0,0,1,1,0}+\frac{1}{8}\,\cL_{0,0,1,1,1}+\frac{1}{768}\,\cL_{0,1,0,0,0}+\frac{1}{24}\,\cL_{0,1,0,0,1}+\frac{1}{32}\,\cL_{0,1,0,1,0}\\
&\qquad+\frac{1}{8}\,\cL_{0,1,0,1,1}+\frac{1}{96}\,\cL_{0,1,1,0,0}+\frac{1}{8}\,\cL_{0,1,1,0,1}+\frac{1}{16}\,\cL_{0,1,1,1,0}+\frac{1}{4}\,\cL_{0,1,1,1,1}-\frac{1}{48}\,\zeta_3\,\cL_{0,1}\, .
\esp\eeq

We do not offer a proof that eqs.~\eqref{R6eqnMHV} and~\eqref{R6eqnNMHV} are valid to all orders in perturbation theory. One may easily check that their expansions through low loop orders, as determined by eqs.~\eqref{eq:g_low_orders} and~\eqref{eq:f_low_orders}, match the known results~\cite{Lipatov2010ad,Dixon2012yy}. It is also straightforward to extend the above calculations to ten loops and confirm that the results are in agreement with those of ref.~\cite{Dixon2012yy}. Moreover, we have verified that the truncated series expansion of eq.~\eqref{R6eqnMHV} as $|w|\to 0$ agrees with that of eq.~\eqref{eq:MHV_MRK_LLA} through 14 loops.

A comparison through such a high loop order is important in order to confirm the absence of multiple zeta values with depth larger than one (hereafter simply ``MZVs''). To see why these MZVs should be absent, consider performing the sum of residues in eq.~\eqref{eq:MHV_MRK_LLA}. Transcendental constants can only arise from the evaluation the $\psi$ function and its derivatives at integer values. The latter are given in terms of rational numbers (Euler-Zagier sums) and ordinary $\zeta$ values. Therefore, it is impossible for the series expansion of eq.~\eqref{eq:MHV_MRK_LLA} to contain MZVs.

On the other hand, we would naively expect MZVs to appear in the series expansion of eq.~\eqref{R6eqnMHV} at 12 loops and beyond.  This expectation is due to the fact that, for high weights, the $y$ alphabet of eq.~\eqref{eq:y_alph} contains MZVs, and, starting at weight 12, these MZVs begin appearing explicitly in the definitions of the SVHPLs. In order for eq.~\eqref{R6eqnMHV} to agree with eq.~\eqref{eq:MHV_MRK_LLA}, all the MZVs must conspire to cancel in the particular linear combination of SVHPLs that appears in~\eqref{R6eqnMHV}. We find that this cancellation indeed occurs, at least through 14 loops. It would be interesting to understand the mechanism of this cancellation, but we postpone this study to future work.
\subsection{Consistency of the MHV and NMHV formulas}
The MHV and NMHV remainder functions are related by the differential equation~\eqref{eq:MHV_NMHV_diffeq},
\beq\label{eq:sec4diffeq}
\ws \frac{\partial}{\partial \ws} R_6^{\textrm{MHV}}|_{\textrm{LLA}}  = w \frac{\partial}{\partial w} R_6^{\textrm{NMHV}}|_{\textrm{LLA}}\, .
\eeq
Recalling that $(w,\ws) = (-z,-\bar{z})$, it is straightforward to use the formulas~\eqref{eq:dLzz} to check that eqs.~\eqref{R6eqnMHV} and~\eqref{R6eqnNMHV} obey this differential equation. To see how this works, consider eq.~\eqref{R6eqnMHV}, which we write as,
\beq
R_6^{\textrm{MHV}}|_{\textrm{LLA}} = \frac{2\pi i}{\log(1-u_1)} \rho\Big[g_0(x_0,x_1)x_0 + g_1(x_0,x_1)x_1\Big]\,,
\eeq
for some functions $g_0(x_0,x_1)$ and $g_1(x_0,x_1)$ which can be easily read off from eq.~\eqref{R6eqnMHV}. The $\ws$ derivative acts on SVHPLs by clipping off the last index and multiplying by $1/\ws$ if that index was an $x_0$ or by $-1/(1+\ws)$ if it was an $x_1$. There are also corrections due to the $y$ alphabet at higher weights. Importantly, $y_0=x_0$, so these corrections only affect the terms with a prefactor $1/(1+\ws)$. This observation allows us to write,
\beq\bsp\label{eq:dwsRMHV}
\ws \frac{\partial}{\partial \ws} R_6^{\textrm{MHV}}|_{\textrm{LLA}} &\;=\;  \frac{2\pi i}{\log(1-u_1)}\rho \Big[g_0(x_0,x_1)- \frac{\ws}{1+\ws}\hat{g}_1(x_0,x_1)\Big]\\
&\;=\; \frac{2\pi i}{\log(1-u_1)}\rho \Big[\frac{1}{1+\ws}g_0(x_0,x_1)+ \frac{1}{1+1/\ws}\Big(g_0(x_0,x_1)-\hat{g}_1(x_0,x_1)\Big)\Big]\,.
\esp\eeq
Due to the complicated expression for $y_1$, it is difficult to obtain an explicit formula for $\hat{g}_1(x_0,x_1)$. Thankfully, we may employ a symmetry argument to avoid calculating it directly. Referring to eq.~\eqref{eq:MHV_MRK_LLA}, $R_6^{\textrm{MHV}}|_{\textrm{LLA}}$ has manifest symmetry under inversion $(w,\ws) \leftrightarrow (1/w,1/\ws)$, or, equivalently, $(\nu,n) \leftrightarrow (-\nu,-n)$. The differential operator $\ws\,\partial_{\ws}$ flips the parity, so eq.~\eqref{eq:dwsRMHV} should be odd under inversion. Since the two rational prefactors on the second line of eq.~\eqref{eq:dwsRMHV} map into one another under inversion, we can infer that their coefficients must be related\footnote{$\rho$ does not generate any rational functions which might allow these terms to mix together.},
\beq
g_0\left(\frac{1}{w},\frac{1}{\ws}\right) = -g_0(w,\ws)+\hat{g}_1(w,\ws)\, ,
\eeq
where $g_0(w,\ws) = \rho(g_0(x_0,x_1))$ and $\hat{g}_1(w,\ws) = \rho(\hat{g}_1(x_0,x_1))$. It is easy to check that this identity is satisfied for low loop orders\footnote{A general proof would be tantamount to showing that eq.~\eqref{R6eqnMHV} is symmetric under inversion. The latter seems to require another intricate cancellation of multiple zeta values. We postpone this investigation to future work.}.

Using these symmetry properties, we can write,
\beq\bsp\label{eq:dwsMHVg0}
\ws \frac{\partial}{\partial \ws} R_6^{\textrm{MHV}}|_{\textrm{LLA}}
&\;=\; \frac{2\pi i}{\log(1-u_1)}\frac{1}{1+\ws}\rho \Big[g_0(x_0,x_1) \Big]\; - \; \bigg\{(w,\ws) \leftrightarrow \left(\frac{1}{w},\frac{1}{\ws}\right)\bigg\}\,.
\esp\eeq
Turning to the right-hand side of eq.~\eqref{eq:sec4diffeq}, we observe that the differential operator $w\,\partial_w$ acts on eq.~\eqref{R6eqnNMHV} by removing the leading $x_0$ and flipping the sign of the second term,
\beq\label{eq:dwNMHV}
w \frac{\partial}{\partial w} R_6^{\textrm{NMHV}}|_{\textrm{LLA}} =  \frac{2\pi i}{\log(1-u_1)}\,\frac{1}{1+\ws}\,\rho\Big[ \mathcal{X}\cZ^{\textrm{NMHV}}\Big] \; - \; \bigg\{(w,\ws) \leftrightarrow \left(\frac{1}{w},\frac{1}{\ws}\right)\bigg\}\,.
\eeq
Comparing eq.~\eqref{eq:dwsMHVg0} and eq.~\eqref{eq:dwNMHV}, we see that eq.~\eqref{eq:sec4diffeq} is satisfied if $g_0(x_0,x_1)= \mathcal{X}\cZ^{\textrm{NMHV}}$. To verify that this is true, we must extract $g_0(x_0,x_1)$ from $R_6^{\textrm{MHV}}|_{\textrm{LLA}}$. To this end, collect all terms in the argument of $\rho$ with at least one trailing $x_0$ and remove that $x_0$. This procedure gives,
\beq\bsp
g_0(x_0,x_1) &\; =\; \frac{1}{2} \mathcal{X} \sum_{k=2}^{\infty}\left(x_1\, \sum_{n=0}^{k-2}(-1)^n x_0^{k-n-2}\sum_{m=0}^{n} \, \frac{2^{2m-k+1}}{(k-m-1)!} \,\mathfrak{Z}(n,m)\right)\eta^k\\
&\;=\;\mathcal{X}\cZ^{\textrm{NMHV}}\, ,
\esp
\eeq
so we conclude that eq.~\eqref{eq:sec4diffeq} is indeed satisfied.
%%%%%%%%%%%%%%%%%%%%%%%%%%%%%%%%%%
%%%%%%%%%%%%%%%%%%%%%%%%%%%%%%%%%%
\section{Collinear limit}
\label{sec:coll}
In the previous section, we proposed an all-orders formula for the MHV and NMHV remainder functions in MRK. The expressions are effectively functions of two variables, $w$ and $\ws$. The single-valuedness condition allows for these functions to be expressed in a compact way, but the result is still somewhat difficult to manipulate. 

In this section, we study a simpler kinematical configuration: the collinear corner of MRK phase space. To reach this configuration, we begin in multi-Regge kinematics and then take legs 1 and 6 to be nearly collinear. In terms of the cross ratios $u_i$, this limit is
\beq
1-u_1,\, u_2,\, u_3 \sim 0\,,\qquad  x\equiv {u_2\over1-u_1} = \cO(1)\,, \qquad y\equiv {u_3\over1-u_1} \sim 0\,,
\eeq
or, in terms of the $(w,\ws)$ variables, it is equivalent to,
\beq
1-u_1\sim~0\,,\qquad |w|\sim 0\,,\qquad w\sim w^*\,.
\eeq

As we approach the collinear limit, the remainder function can be expanded in powers of $w$, $\ws$, and $\log|w|$. The leading power-law behavior is proportional to $(w+\ws)$. Neglecting terms that are suppressed by further powers of $|w|$, the result is effectively a function of a single variable, $\xi=\eta\log|w|=a\log(1-u_1)\log|w|$, and is simple enough to be computed explicitly, as we show in the following subsections.
%%%%
\subsection{MHV}
In the MHV helicity configuration, the remainder function is symmetric under conjugation $w\leftrightarrow \ws$. It also vanishes in the strict collinear limit. These conditions suggest a convenient form for the expansion in the near-collinear limit,
\beq\label{eq:R6_MHV_coll}
R_6^{\textrm{MHV}}|_{\textrm{LLA, coll.}} \;=\; \frac{2 \pi i}{\log (1-u_1)}(w+\ws) \sum_{k=0}^{\infty} \eta^{k+1} \, r_{k}^{\textrm{MHV}}\!\big(\eta \log|w|\big)\,,
\eeq
for some functions $r_{k}^{\textrm{MHV}}$ that are analytic in a neighborhood of the origin. We have neglected further power-suppressed terms, $i.e.$ terms quadratic or higher in $w$ or $\ws$. The index $k$ labels the degree to which $r_{k}^{\textrm{MHV}}$ is subleading in $\log|w|$. For example, the leading logarithms are collected in $r_{0}^{\textrm{MHV}}$, the next-to-leading logarithms are collected in $r_{1}^{\textrm{MHV}}$, etc.

Starting from eq.~\eqref{R6eqnMHV}, it is possible to obtain an explicit formula for $r_{k}^{\textrm{MHV}}$. To begin, we note that it is sufficient to restrict our attention to the terms proportional to $w$ --- the conjugation symmetry guarantees that they are equal to the terms proportional to $\ws$. The main observation is that only a subset of terms in eq.~\eqref{R6eqnMHV} contribute to the power series expansion at order $w$. It turns out that the relevant subset is simply the set of SVHPLs with a single $x_1$ in the weight vector. Roughly speaking, each additional $x_1$ implies another integration by $1/(1+w)$, which increases the leading power by one.

The equivalent statement is not true for $\ws$, i.e. SVHPLs with an arbitrary number of $x_1$'s contribute to the power series expansion at order $\ws$. This asymmetry can be traced to the differences between the $x$ and $y$ alphabets: referring to eq.~\eqref{eq:LXY0}, the $x$ alphabet indexes the HPLs with argument $w$ and the $y$ alphabet indexes the HPLs with argument $\ws$. 

We are therefore led to consider the terms in eq. \eqref{R6eqnMHV} with exactly one $x_1$. Eq.~\eqref{eq:XZZ} shows that these terms may be obtained by dropping all $x_1$'s from $\mathcal{X}$,
\beq
\label{R6eqn_coll}
R_6^{\textrm{MHV}}|_{\textrm{LLA, coll.}} = \frac{2\pi i}{\log(1-u_1)}\,\rho\Big( e^{\frac{1}{2}x_0 \eta}\cZ^{\textrm{MHV}}- \frac{1}{2}\,x_1 \eta\Big)\, .
\eeq
Since no $\zeta$ terms appear in SVHPLs with a single $x_1$, it is straightforward to express them in terms of HPLS,
\beq\label{SVHPLasHPL}
\cL_{x_0^n x_1 x_0^m} = \sum_{j=0}^{n} \frac{1}{j!}H_{x_0}^{j}\Hb_{x_0^mx_1x_0^{n-j}} \; + \; \sum_{j=0}^{m} \frac{1}{j!}H_{x_0^n x_1x_0^{m-j}}\Hb_{x_0}^{j}\, .
\eeq
Here we have simplified the notation by defining $H_m\equiv H_m(-w)$ and $\Hb_m=H_m(-\ws)$. Next, we recall eq.~\eqref{eq:shuff_x1}, in which we used the shuffle algebra to expose the explicit logarithms,
\beq
H_{x_0^n x_1 x_0^m} = \sum_{j=0}^{m}\frac{(-1)^{j}}{(m-j)!}\binom{n+j}{j}H_{x_0}^{m-j} H_{x_0^{n+j}x_1}\,.
\eeq
Finally, eqs.~\eqref{eq:HPL_series} and~\eqref{eq:Euler-Zagier} implies that the series expansions for small $w$ have leading term,
\beq\label{eq:H01exp}
H_{x_0^k x_1}(-w) = -w + \cO(w^2) \, .
\eeq

Combining eqs.~\eqref{R6eqn_coll}-\eqref{eq:H01exp} and applying some hypergeometric function identities, we arrive at an explicit formula for $r_{k}^{\textrm{MHV}}$,
\beq\label{eq:rkMHV}
r_{k}^{\textrm{MHV}}(x) =\frac{1}{2}\,\delta_{0,k}+ \sum_{n=0}^{k}\sum_{m=0}^{n} \sum_{j=k-m}^{2k-n-m} \frac{(-2)^{2m+j-k-1}}{(m+j-k)!}\,\mathfrak{Z}(n,m)\,x^{m-k+j/2}\,P_{j}^{(k-j-n, k-j-m)}\big(0\big)\,I_{j}\big(2\sqrt{x}\,\big)\,.
\eeq
In this expression, the $I_j$ are modified Bessel functions and the $P_j^{(a, b)}$ are Jacobi polynomials, which can be defined for non-negative integers $j$ by the generating function,
\beq\label{eq:jacobiP}
%P_j^{(a, b)}(x) = \frac{\Gamma(1+a+j)}{\Gamma(1+a)\Gamma(1+j)}\,\phantom{}_2F_1\left(-j,1+a+b+j;a+1,\frac{1-x}{2}\right)\, .
\sum_{j=0}^{\infty} P_j^{(a,b)}(z)\,t^j = 2^{a+b}\,\Big(1-t+\sqrt{t^2-2tz+1}\Big)^{-a}\Big(1+t+\sqrt{t^2-2tz+1}\Big)^{-b}\Big(\sqrt{t^2-2tz+1}\Big)^{-1}\,.
\eeq

It is easy to extract the first few terms,
\begin{align}
r_{0}^{\textrm{MHV}}(x) &\;=\; \frac{1}{2}\left[1-I_0\left(2\sqrt{x}\right)\right]\, ,\nonumber\\
r_{1}^{\textrm{MHV}}(x) &\;=\; -\frac{1}{4}\,I_2\left(2\sqrt{x}\right)\,,\\
r_{2}^{\textrm{MHV}}(x) &\;=\; \frac{1}{4x}\,I_2\left(2\sqrt{x}\right) -\frac{1}{16}\,I_4\left(2\sqrt{x}\right)\nonumber \, .
\end{align}
The leading term, $r_{0}^{\textrm{MHV}}$, corresponds to the double-leading-logarithmic approximation (DLLA) of ref.~\cite{Bartels2011xy},
\beq
R_6^{\textrm{MHV}}|_{\textrm{DLLA}} \;=\;i\pi\,a\,(w+\ws)\,\left[1-I_0\left(2\sqrt{\eta\log|w|}\right)\right]\,,
\eeq
and is in agreement with the results of that reference.

Only for $k>2$ do $\zeta$ values begin to appear in $r_{k}^{\textrm{MHV}}$. Moreover, modified Bessel functions with odd indices only appear in the $\zeta$-containing terms. To see this, notice that the $\zeta$-free terms of eq.~\eqref{eq:rkMHV} arise from the boundary of the sum with $n=m=0$, in which case $a=b=k-j$ in eq.~\eqref{eq:jacobiP}. When $a=b$, $P^{a,b}_j(0)=0$ for odd $j$ since eq.~\eqref{eq:jacobiP} reduces to a function of $t^2$ in this case. It follows that the $\zeta$-free pieces of $r_{k}^{\textrm{MHV}}$ have no modified Bessel functions with odd indices.

Equations~\eqref{eq:R6_MHV_coll} and~\eqref{eq:rkMHV} provide an explicit formula for the six-point remainder function in the near-collinear limit of the LL approximation of MRK. If the sum in eq.~\eqref{eq:R6_MHV_coll} converges sufficiently quickly, then it should be possible to evaluate the function numerically by truncating the sum at a finite value of $k$, $k_\textrm{max}$. A numerical analysis indicates that for $|w|<1$ and $\eta \lesssim 20$, $k_\textrm{max} \simeq 100$ is adequate to ensure convergence.

\begin{figure}[t]
\centering
\includegraphics[height=100mm]{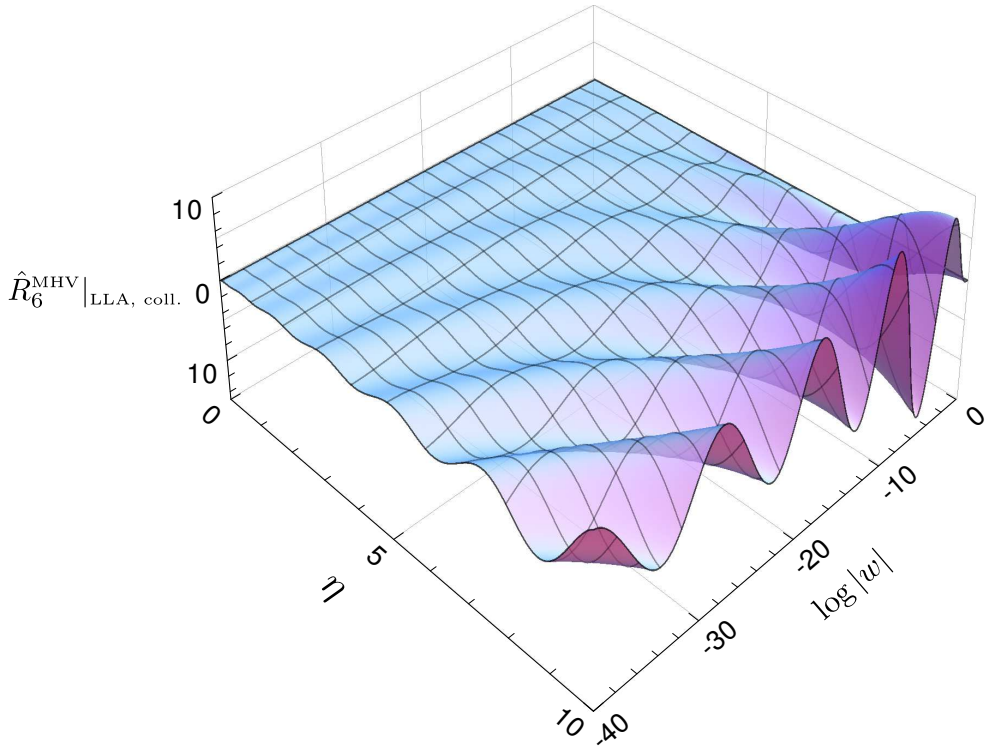}
\caption{The MHV remainder function in the near-collinear limit of the LL approximation of MRK. It has been rescaled by an exponential damping factor. See eq.~\eqref{eq:damping}.}
\label{fig:R6_MHV_coll}
\end{figure}

The numerical analysis also indicates that $R_6^{\textrm{MHV}}|_{\textrm{LLA, coll.}}$ increases exponentially as a function of $\eta$, and that the extent of this increase depends strongly on the value of $\log|w|$. We find empirically that the rescaled function
\beq\label{eq:damping}
\hat{R}_6^{\textrm{MHV}}|_{\textrm{LLA, coll.}} = \exp\bigg(\!\!-\!\frac{\eta}{\sqrt[4]{-\log|w|}}\bigg)\,\frac{\log(1-u_1)}{2\pi i \,(w+\ws)}\,R_6^{\textrm{MHV}}|_{\textrm{LLA, coll.}}
\eeq
attains reasonable uniformity in the region $0<\eta<10$ and $-40<\log|w|<0$. This particular rescaling carries no special significance, as alternatives are possible and may be more appropriate in different regions. In eq.~\eqref{eq:damping} we have also divided by the overall prefactor of eq.~\eqref{eq:R6_MHV_coll} so that $\hat{R}_6^{\textrm{MHV}}|_{\textrm{LLA, coll.}}$ is truly a function of the two variables $\eta$ and $\log|w|$. The results are displayed in Figure~\ref{fig:R6_MHV_coll}.

%%%%%%%%%%
\subsection{NMHV}
A similar analysis can be performed for the NMHV helicity configuration. The situation is slightly more complicated in this case because the NMHV remainder function is not symmetric under conjugation $w\leftrightarrow\ws$. One consequence is that its expansion in the collinear limit requires two sequences of functions, which we choose to parameterize by $r_{k}^{\textrm{NMHV}}$ and $\tilde{r}_{k}^{\textrm{NMHV}}$,
\beq\bsp
R_6^{\textrm{NMHV}}|_{\textrm{LLA, coll.}} &\;=\; \frac{2 \pi i}{\log (1-u_1)}\Bigg[(w+\ws) \sum_{k=0}^{\infty} \eta^{k+2} \, r_{k}^{\textrm{NMHV}}\!\big(\eta \log|w|\big) \\
&\hspace{3.75cm}+ \ws \sum_{k=0}^{\infty} \eta^{k} \, \tilde{r}_{k}^{\textrm{NMHV}}\!\big(\eta \log|w|\big)\Bigg]\, .
\esp\eeq
Contributions to the power series at order $w$ arise from the first term of eq.~\eqref{R6eqnNMHV} (the second term has an overall factor of $\ws$), and, as in the MHV case, only from the subset of SVHPLs with a single $x_1$ in the weight vector. It is therefore possible to reuse eqs.~\eqref{SVHPLasHPL}-\eqref{eq:H01exp} and obtain an explicit formula for the coefficient of $w$, $r_{k}^{\textrm{NMHV}}$. The result is,
\beq\label{eq:rkNMHV}
r_{k}^{\textrm{NMHV}}(x) = \sum_{n=0}^{k}\sum_{m=0}^{n} \sum_{j=k-m}^{2k-n-m} \frac{(-2)^{2m+j-k}}{(m+j-k)!}\,\mathfrak{Z}(n,m)\,x^{m-k+(j-1)/2}\,P_{j+2}^{(k-j-n-1, k-j-m-1)}\big(0\big)\,I_{j+1}\big(2\sqrt{x}\,\big)\, .
\eeq
The first few terms are
\beq\bsp
r_{0}^{\textrm{NMHV}}(x) &\;=\; -\frac{1}{4 \sqrt{x}} I_1(2\sqrt{x})\, ,\\
r_{1}^{\textrm{NMHV}}(x) &\;=\; -\frac{1}{8\sqrt{x}}I_3(2\sqrt{x})\, ,\\
r_{2}^{\textrm{NMHV}}(x) &\;=\;\frac{3}{16x^{3/2}}I_3(2\sqrt{x})-\frac{1}{32\sqrt{x}}I_5(2\sqrt{x})\, .
\esp\eeq

As previously mentioned, it is not so straightforward to extract the coefficient of $\ws$ in this way. We can instead make progress by exploiting the differential equation~\eqref{eq:MHV_NMHV_diffeq}. In terms of the functions $r_{k}^{\textrm{MHV}}$, $r_{k}^{\textrm{NMHV}}$, and $\tilde{r}_{k}^{\textrm{NMHV}}$, the equations read,
\beq\bsp
\partial_x r_{k}^{\textrm{MHV}}(x) &\; =\; 2\,r_{k}^{\textrm{NMHV}}(x)+\partial_x r_{k-1}^{\textrm{NMHV}}(x)\\
\partial_x \tilde{r}_{k}^{\textrm{NMHV}}(x) &\;=\; 2\, r_{k}^{\textrm{MHV}}(x)+ 2\,r_{k-1}^{\textrm{NMHV}}(x)\, .
\esp\eeq
The first of these equations is automatically satisfied and confirms the consistency of eq.~\eqref{eq:rkMHV} and eq.~\eqref{eq:rkNMHV}.
The second equation determines $\tilde{r}_{k}^{\textrm{NMHV}}$ up to a constant of integration which can be determined by examining the $n=-1$ term of eq.~\eqref{eq:NMHV_MRK}. The solution is,
\beq\label{eq:rtildekNMHV}
\tilde{r}_{k}^{\textrm{NMHV}}(x) = x\, \delta_{0,k}-\sum_{n=0}^{k}\sum_{m=0}^{n} \sum_{j=k-m}^{2k-n-m} \frac{(-2)^{2m+j-k}}{(m+j-k)!}\,\mathfrak{Z}(n,m)\,x^{m-k+(j+1)/2}\,P_{j}^{(k-j-n, k-j-m)}\big(0\big)\,I_{j-1}\big(2\sqrt{x}\,\big)\, .
\eeq
The first few terms are
\beq\bsp
\tilde{r}_{0}^{\textrm{NMHV}}(x) &\;=\; x-\sqrt{x}\,I_1\left(2\sqrt{x}\right)\, ,\\
\tilde{r}_{1}^{\textrm{NMHV}}(x) &\;=\; -\frac{1}{2}\sqrt{x}\,I_1\left(2\sqrt{x}\right)\, ,\\
\tilde{r}_{2}^{\textrm{NMHV}}(x) &\;=\; \frac{1}{2\sqrt{x}}\,I_1\left(2\sqrt{x}\right) -\frac{1}{8}\sqrt{x}\,I_3\left(2\sqrt{x}\right)\,.
\esp\eeq
Modified Bessel functions with even indices only appear in the $\zeta$-containing terms of $r_{k}^{\textrm{NMHV}}$ and $\tilde{r}_{k}^{\textrm{NMHV}}$. The explanation of this fact is the same as in the MHV case, except that the parity is flipped due to the shifts of the indices of the modified Bessel functions in eq.~\eqref{eq:rkNMHV} and eq.~\eqref{eq:rtildekNMHV}.
\subsection{The real part of the MHV remainder function in NLLA}
As described in Section~\ref{sec:MRK}, the real part of the MHV remainder function in NLLA is related to its imaginary part in LLA. In the collinear limit, the relation~\eqref{eq:ReNLLA} may be written as,
\beq\bsp\label{eq:ReNLLAcoll}
\textrm{Re}\left(R_6^{\textrm{MHV}}|_{\textrm{NLLA, coll.}}\right) & \;=\; \frac{2\pi i}{\log(1-u_1)}\left(\frac{1}{2}\,\eta^2\,\frac{\partial}{\partial \eta}\,\frac{1}{\eta} - \frac{1}{2}\,\eta\log|w|\right)\,R_6^{\textrm{MHV}}|_{\textrm{LLA, coll.}} \\
&\quad   - \frac{\pi^2}{\log^2(1-u_1)}\eta^2 \log|w|\, .
\esp\eeq
Since $R_6^{\textrm{MHV}}|_{\textrm{LLA}}$ vanishes like $(w+\ws)$ in the strict collinear limit, the quadratic term $(R_6^{\textrm{MHV}}|_{\textrm{LLA}})^2$ in eq.~\eqref{eq:ReNLLA} only contributes to further power-suppressed terms in the near-collinear limit and is therefore omitted from eq.~\eqref{eq:ReNLLAcoll}\footnote{As a consequence, eq.~\eqref{eq:ReNLLAcoll} does not depend on the conventions used to define $R$, i.e. the equation is equally valid if $R$ is replaced by $\exp(R)$.}. We may write eq.~\eqref{eq:ReNLLAcoll} as,
\beq
\textrm{Re}\left(R_6^{\textrm{MHV}}|_{\textrm{NLLA, coll.}}\right) = -\frac{4 \pi^2}{\log^2(1-u_1)}(w+\ws)\sum_{k=0}^{\infty}\eta^{k+1}q_{k}\big(\eta \log|w|\big)\, ,
\eeq
where,
\beq
q_k(x) = \frac{1}{4}\,x\,\delta_{0,k}+\frac{1}{2}\,(k-x)\,r_{k}^{\textrm{MHV}}\!\big(x\big) + \frac{1}{2}\,x \partial_x r_{k}^{\textrm{MHV}}\!\big(x\big)\, .
\eeq
The leading term, $q_0$, corresponds to the real part of the next-to-double-leading-logarithmic approximation (NDLLA) of ref.~\cite{Bartels2011xy}. Our results agree\footnote{The agreement requires a few typos to be corrected in eq.~(A.16) of ref.~\cite{Bartels2011xy}.} with that reference and read,
\beq
\textrm{Re}\left(R_6^{\textrm{MHV}}|_{\textrm{NDLLA}}\right) \;=\;\frac{\pi^2\,(w+\ws)\,\eta}{\log^2(1-u_1)}\,\left[-\eta\log|w| I_0\left(2\sqrt{\eta\log|w|}\right) + \sqrt{\eta\log|w|}\,I_1\left(2\sqrt{\eta\log|w|}\right)\right]\,.
\eeq
\section{Conclusions}
\label{sec:conclusion}
In this article, we studied the six-point amplitude of planar $\cN=4$ super-Yang-Mills theory in the leading-logarithmic approximation of multi-Regge kinematics. In this limit, the remainder function 
assumes a particularly simple form, which we exposed to all loop orders in terms of the single-valued harmonic polylogarithms introduced by Brown. The SVHPLs provide a natural basis of functions for the remainder function in MRK because the single-valuedness condition maps nicely onto a physical constraint imposed by unitarity. Previously, these functions had been used to calculate the remainder function in LLA through ten loops. In this work, we extended these results to all loop orders.

In MRK, the tree amplitudes in the MHV and NMHV helicity configurations are identical. This observation motivates the definition of an NMHV remainder function in analogy with the MHV case. We examined both remainder functions in this article, and proposed all-order formulas for each case. In fact, these formulas are related: as described in ref.~\cite{Lipatov2012gk}, the two remainder functions are linked by a simple differential equation. We employed this differential equation to verify the consistency of our results.

We also investigated the behavior of our formulas in the near-collinear limit of MRK. The additional large logarithms that arise in this limit impose a hierarchical organization of the resulting expansions. We derived explicit all-orders expressions for the terms of this logarithmic expansion. The results are given in terms of modified Bessel functions.

We did not provide a proof of the all-orders result, but we verified that it agrees through 14 loops with an integral formula of Lipatov and Prygarin. The agreement of these formulas at 12 loops and beyond requires an intricate cancellation of multiple zeta values. It would be interesting to understand the mechanism of this cancellation. There are several other potential directions for future research. For example, in refs.~\cite{Alday2010ku,Gaiotto2010fk,Gaiotto2011dt}, Alday, Gaiotto, Maldacena, Sever, and Vieira performed an OPE analysis of hexagonal Wilson loops which in principle should provide additional cross-checks of our results. It should also be possible to study the all-orders formula as a function of the coupling and, in particular, to examine its strong-coupling expansion. We have begun this study in the collinear limit and presented our initial results in Figure~\ref{fig:R6_MHV_coll}. A first attempt to compare the six-point remainder function in MRK at strong and weak coupling was made by Bartels, Kotanski, and Schomerus~\cite{Bartels2010ej}. Further analysis of our all-orders formula should allow for an important comparison with this string-theoretic calculation.

\section*{Acknowledgments}
I am grateful to Lance Dixon and Claude Duhr for many helpful discussions and comments on the manuscript.
This research was supported by the US Department of Energy under
contract DE--AC02--76SF00515.

\end{document}